# Association of vaccine-induced or hybrid immunity with COVID-19-related mortality during the Omicron wave - a retrospective observational study in elderly Bavarians

Maximilian Weigert, MSc (Maximilian.Weigert@stat.uni-muenchen.de)[1]

Andreas Beyerlein, PhD (Andreas.Beyerlein@lgl.bayern.de)[2]

Katharina Katz, PhD (Katharina.Katz@lgl.bayern.de)[2]

Rickmer Schulte, BSc (rickmer.schulte@stablab.stat.uni-muenchen.de)[1]

Wolfgang Hartl, MD (whartl@med.uni-muenchen.de)[3] [#]

Helmut Küchenhoff, PhD (kuechenhoff@stat.uni-muenchen.de)[1] [#]

[1]) Statistical Consulting Unit StaBLab, Department of Statistics, Ludwig-Maximilians-Universität Munich, Germany
[2]) Bavarian Health and Food Safety Authority, Oberschleißheim, Germany
[3]) Division of Surgical Critical Care, Department of General, Visceral, and Transplantation Surgery, University Medical Center, Campus Grosshadern, Ludwig-Maximilians-Universität Munich, Germany

[#]) Contributed equally

# Abstract

**Background:** There is a lack of population-based studies on the effectiveness and durability of the SARS-CoV-2-induced immune protection during the Omicron wave.

**Methods:** This retrospective study included 470 159 cases aged 60 years or older, who tested positive for SARS-CoV-2 between January 1 and June 30, 2022 in Bavaria, Germany. We examined time to death, measured from the earliest date of reporting/symptom onset until day 60. Cox models were used to estimate adjusted hazard ratios (HR) for sex, age, earliest calendar date of documented infection, and level of vaccine-induced or hybrid immunity (LI).

**Results:** We observed 3836 deaths (case fatality rate 0.82%). Risk of death was significantly lower in all LIs than in unvaccinated cases (adjusted HR for a full primary LI achieved within the last six months: 0.30, 95% confidence interval (CI) 0.23-0.39; after six months: 0.46, 95% CI 0.35-0.60). Boosting further decreased the risk of death (within the last three months: HR 0.17, 95% CI 0.15-0.20; after three months: HR 0.25, 95% CI 0.21-0.29).

**Conclusion:** In elderly Bavarians, an increasing LI was associated with an increasing protection against death during the Omicron wave. Protection, however, may decrease to some extent after several months.

**Introduction**

A major epidemic of the SARS-CoV-2 Omicron (B.1.1.529) variant sublineage BA.1 / BA.2 began in Bavaria, Germany, in early January 2022. Details on simultaneous infection protection measures and course of the Omicron wave are presented in the online supplement. Despite ongoing restrictions, this subvariant lead to a sharp rise of the weekly number of reported SARS-CoV-2 infections and hospital / intensive care unit (ICU) admissions. At the height of the Omicron endemic, however, occupancy of Bavarian ICU beds with COVID-19 patients was only one fifth of the peak observed during the Delta wave (**1**) reflecting - together with a more than 50% lower death toll (**2**) - a milder course of disease.

The reasons for this attenuated Omicron-associated morbidity and mortality are still unclear. It has been argued that the COVID-19 vaccines which were available in early 2022 had been developed on older variants of concern, and therefore their effectiveness and durability may be reduced with respect to SARS-CoV-2 Omicron sub-lineages BA.1 and BA.2 (**3 -7**).

Knowledge on the effectiveness of vaccine-induced or hybrid immunity against mortality during the Omicron wave has been mainly derived from large scale studies in populations outside Germany (**8**). Findings may not be generalized to all countries because of national specificities including levels of immunity. In Germany, recent studies have only assessed vaccine effectiveness, but not effectiveness of a combined vaccine-induced and hybrid immunity (**9, 10**). Furthermore, these studies only used methods for aggregated data (**9**), or only included hospitalized patients (**10**).

Here, we analyzed the data of all recorded SARS-Cov-2 infections from January to June 2022 in elderly Bavarians to better understand the associations between vaccine-induced or hybrid immunity, and the rate of death during the Omicron wave.

**Methods**

*Study design and population*

In this retrospective cohort study, we analyzed RT-PCR confirmed Bavarian cases, which had been officially registered with a SARS-CoV-2 infection between January 1 (date of rebound in daily SARS-CoV-2 infections in Bavaria (**11**)) and June 30, 2022, when relative frequencies of Omicron BA.1 and Omicron BA.2 infections had fallen below 10%, and the Omicron BA.5 variant was dominating (**12**).

This was a secondary analysis of anonymized data from the Bavarian database for mandatorily notifiable infectious diseases according to the German Infection Protection Act (Infektionsschutzgesetz, IfSG) managed by the Bavarian infection surveillance system (Bavarian Health and Food Safety Authority, in German: Landesamt für Gesundheit und Lebensmittelsicherheit, LGL). The IfSG obliges all testing institutions to inform the public health departments of positive RT-PCR test results and of associated demographic variables. For each case, the local public health departments receive personal information such as name and address, and are advised to record further information such as (but not limited to) sex, age, vaccination status, date of last vaccination, date of symptom onset, SARS-CoV-2 infection history (first infection or re-infection) and date of potentially COVID-19-related death. The LGL collects the pseudonymized data of all SARS-CoV-2 cases (including age, gender and date of preceding infections, symptom onset and of reporting) recorded by the local Bavarian public health departments on a daily basis.

Between January and June 2022, a total of 3 846 297 Bavarian cases had been reported on this basis. Since the largest increase in mortality risk had been observed in COVID-19 patients older than 59 years (**13, 14**), and since more than 95% of deaths had occurred in German patients older than 59 years during the Omicron wave (**15**), we restricted our analyses to this age group.

Final update and data extraction was on September 7, 2022. The data analysis was approved by the institutional review board of the LMU Munich (project # 22-0429).

*Data*

We obtained death registration data (including date of death) from the LGL for all registered cases which were known to have died after a test had revealed a preceding SARS-CoV-2 infection. COVID-19 could have either been the direct (leading) or indirect cause of death; the cause of death could have also been classified as unknown or undetermined, if COVID-19 could not be excluded as a potential cause of death. Details on the classification of causes of death, and on procedures used to handle implausible/missing calendar dates of death are presented in the online supplement. Primary outcome was survival time.

Due to personal data protection regulations, individual age information was converted into 5-year age categories at data extraction. The earliest date of documented infection was set to either the calendar date of reporting, or of symptom onset, whatever came first.

Levels of immunity were classified as follows: unknown, not vaccinated, incomplete primary level of immunity (one vaccination without a preceding SARS-CoV-2 infection), full primary level of immunity (two vaccinations or one vaccination combined with a preceding SARS-CoV-2 infection) more or less than six months before the earliest date of documented infection, boosted level of immunity (three or four vaccinations, or two vaccinations combined with a preceding SARS-CoV-2 infection more or less than three months before the earliest date of documented infection). Cases with a previous SARS-CoV-2 infection, but without any vaccination, were considered to be unvaccinated. Details on vaccines, on the definition of the level of immunity, and on timing of vaccine authorizations are presented in the online supplement.

*Statistical analysis*

Cox proportional hazard regression (**16**) was used to estimate mutually adjusted hazard ratios describing risks of COVID-19-related death given the level of immunity, sex, age and earliest date of documented infection. For categorical covariates, fixed effects were estimated. The time of the earliest date of documented infection, and age were included into our model using penalized splines (**17**) with four degrees of freedom. Thus, we controlled for potential time-varying confounders (**18**). Estimates for adjusted absolute risk reduction and effectiveness were derived from the fitted Cox model. Further details on statistical methods including a subgroup analysis in cases older than 79 years, and sensitivity analyses are presented in the online supplement.

**Results**

Of all 3 846 297 RT-PCR primary positive cases in Bavaria between January 1 and June 30, 2022, 3 344 229 were excluded because of an age <60 years, an earliest date of documented infection before January 1, 2022 or missing demographic data (age, sex) (Figure 1). Of the remaining 502 068 cases, 31 909 (6.4%) had to be excluded because of an unknown outcome. This fraction peaked at the height of the omicron wave, but remained < 9% throughout the study (calendar week 1 to 26, Figure S1). Finally, 470 159 cases could be analyzed. Most of them (168 817, 35.9%) were between 60 and 64 years old, 248 312 (52.8%) were female and 48 (0.01%) diverse; 94 457 cases (20.1%) were older than 80 years at the earliest date of documented infection. 1.7% of infected cases had already had a documented preceding SARS-CoV-2 infection.

*Overall mortality rates*

Of eligible cases, 3 836 died until the end of the observation period (60-day case fatality rate, CFR 0.82%). Most of the deaths (25.0%) occurred in those aged 90 years or older. Median time until death was 8 days (IQR 4 – 14 days).

Among deceased cases, COVID-19 was the leading cause of death in 2 663 (69.4%); in 871 cases (22.7%), a SARS-CoV-2 infection was not directly linked to death, and in 302 cases (7.9%), cause of death was unidentifiable or was not determined. Despite a considerable daily variation in death toll, the relative proportion of these categories remained fairly stable during the Omicron wave (Figure S2).

*Level of immunity*

Demographic characteristics and outcomes of cases according to the level of immunity are presented in Table 1. For the vast majority of our cases (n = 343 018), level of immunity was unknown (73.0%). The relative portion of cases with an unknown level remained largely stable between February and June 2022 (60% - 80%) (Figure S3).

Among cases, of which the level of immunity had been known, 79 083 (62.2%) had a boosted level of immunity, 15 652 a full primary level of immunity (12.3%) and 19 561 had not been vaccinated (15.4%).

Kaplan-Meier graphs for survival probabilities according to the level of immunity are presented in Figure 2. Survival probability increased by the level of immunity; time since achieving a certain level of immunity appeared to be related to outcome. Survival probability of cases, of which the level of immunity was unknown, was similar to that of cases which had achieved a boosted level of immunity more than three months before the earliest date of documented infection.

*Multivariable analysis: Independent variables associated with the risk of death*

As none of the diverse cases had died, these cases were not further analyzed. In multivariable analysis with mutual adjustment, risk of death was significantly higher in male cases (adjusted hazard ratio, HR 1.70, 95% confidence interval, CI 1.59-1.81) compared to female cases. We identified a significant non-linear association of the risk of death with both age and earliest date of documented infection (Figures 3a and 3b).

Compared to cases which had been unvaccinated, the adjusted risk of death for an incomplete primary level of immunity was significantly lower (HR 0.42, 95% CI 0.35-0.50, effectiveness 56.7%, 95% CI 48.9-63.5%; absolute risk reduction 1.4%, 95% CI 1.2%-1.8%) (Figure 4). Risk of death further decreased with an increasing level of immunity, and was lowest in cases which had achieved their boosted level of immunity less than three months before the earliest date of documented infection (HR 0.17, 95% CI 0.15–0.20, effectiveness 81.9%, 95% CI 78.8-84.7%; absolute risk reduction 2.1%, 95% CI 1.9-2.4%).

The HR of cases with an incomplete primary level of immunity was in the range of that calculated for cases which had achieved a full primary level of immunity more than six months before the earliest date of documented infection. Risk of death in cases of which level of immunity had been unknown (HR 0.28, 95% CI 0.25-0.31, effectiveness 71.0%, 95% CI 68.0-73.7%; absolute risk reduction 1.8%, 95% CI 1.6%-2.1%) corresponded to the risk calculated for cases which had achieved a full primary level of immunity less than six months before the earliest date of documented infection.

There was evidence for some waning of protection provided by a certain level of immunity. With a full primary level of immunity, HR increased after six months from 0.30 (95% CI 0.23-0.39) to 0.46 (95% CI 0.35-0.60), effectiveness decreased from 69.1% (95% CI 61.1-76.1%) to 52.8% (95% CI 40.2-64.9%), and absolute risk reduction from 1.8% (95% CI 1.5-2.1%) to 1.4% (95% CI 1.0-1.7%), respectively.

Corresponding changes were seen with a boosted level of immunity of which the associated HR increased after three months from 0.17 (95% CI 0.15-0.20) to 0.25 (95% CI 0.21-0.29), effectiveness decreased from 81.9% (95% CI 78.8.-84.7%) to 74.5% (95% CI 70.1-78.7%), and absolute risk reduction from 2.1% (95% CI 1.9-2.4%) to 1.9% (95% CI 1.7-2.2%).

Sensitivity analyses and an exclusive analysis of octo- and nonagenarians yielded qualitatively similar findings to the main analysis (see online supplement, supplementary Figures S4 to S11).

**Discussion**

A detailed discussion of mortality, causes of death, demographic predictors of mortality risk (age, gender), and of associations between the earliest date of documented infection and mortality risk can be found in the online supplement.

One key finding of our study was that, in comparison to unvaccinated cases, an increasing level of immunity was associated with a decreasing risk of death from the Omicron variant in Bavaria (including cases older than 79 years); in absolute terms, risk fell from 2.6% to 0.5%, if cases had achieved a boosted level of immunity less than three months before the earliest date of documented infection. If, however, only a full primary level of immunity had already been established more than half a year before that date, risk of death was still appreciable in public health terms (effectiveness only about 53%, absolute risk reduction only 1.4%), and corresponded to that associated with an incomplete primary level of immunity. Even with a boosted level of immunity, however, effectiveness against death (75-82%) was about 15% less than that observed during the Delta wave (**6**). This difference may relate to the type of variant or to the fact that we exclusively studied elderly patients.

A previous symptomatic or asymptomatic SARS-CoV-2 infection may offer high protection against a subsequent Omicron-associated hospitalization and death in unvaccinated and vaccinated individuals (**7, 18, 19-22**). Since, in our data, only 1.7% of elderly infected Bavarians had already had a preceding SARS-CoV-2 infection, it is likely that in our study vaccination status largely determined the associations between level of immunity and survival.

Qualitatively, our results on primary and boosted levels of immunity (including the associated effectiveness) are largely in line with observations from numerous other studies which studied vaccine effectiveness, and of which some also examined elderly patients (**7, 18, 23-31**). It is currently uncertain, however, if, when and to which degree immunity wanes after vaccination or a preceding infection. Concerning the prevention of a severe disease, a study from Brazil, and one older review on the subject were unable to provide clear evidence for a significant waning of full primary immunization during the Omicron wave (**23, 32**). On the other hand, waning after four to six months was suggested in recent studies which examined vaccine effects or effects of a preceding infection on hospitalization/death after an Omicron infection (**33, 29, 34-36**).

Our results suggest that there could be some waning of immunity six months after a full primary level of immunity had been achieved. Effectiveness of a full primary level of immunity decreased from 69% before, to 53% after six months. However, corresponding 95% confidence intervals were slightly overlapping.

Similarly, it is currently also controversial, at which rate effects of booster immunizations on the risk for a severe Omicron disease are declining. Two studies were unable to identify a significant waning up to six months after boosting (**28, 33**), whereas numerous other studies had suggested a faster, but sometimes only a small waning of first or second booster vaccine effect especially in elderly cases (**4, 5, 7, 18, 29-31, 34, 35, 37-39**). The results of our study seem to confirm the latter findings. Three months after boosting, effectiveness fell slightly, but significantly from 82 to 75%.

*Limitations*

As with any retrospective observational analysis, there was a possibility of residual confounding due to several unmeasured covariates (comorbidities, health behavior, socio-economic status, ethnicity, BMI or testing behavior) and of incorrect or absent reporting of important variables (type of vaccine, cause of death, vaccination status). A detailed discussion of the limitations of this study, their individual importance, and their consideration in the context of this study, can be found in the online supplement.

*Strength*

To our knowledge, this is the first German cohort study analyzing the association between individual levels of immunity and survival time during the Omicron wave. By additionally adjusting to the earliest calendar date of documented infection, we were able to adjust for further time-dependent variables improving the validity of our results.

*Conclusion*

Our results demonstrate that a high level of immunity offered a significant protection against COVID-19-related mortality in Bavaria, even in octo- and nonagenarians. Older Bavarian people, however, remain at increased risk of death even with a full primary level of immunity. Since it may wane after six months, they should be prioritized for additional boosters. A boosted immunity did provide further protection, but also this protection becomes somewhat weaker after three months.

**Declarations**

*Ethical approval and consent to participate*

This non-interventional study and the retrospective anonymous data analysis was approved by local institutional review boards which also waived the need for an informed written consent. All applicable data protection regulations were followed to ensure patient confidentiality.

*Consent for publication*

All authors critically reviewed and approved the final submitted version of the manuscript. All authors have agreed both to be personally accountable for the author's own contributions and to ensure that questions related to the accuracy or integrity of any part of the work, even ones in which the author was not personally involved, are appropriately investigated, resolved, and the resolution documented in the literature.The authors of this work take full responsibility for its content.

*Availability of data and materials*

The analysis code deposited at time of publication is available in an open source GitHub repository at https://github.com/MaxWeigert/COVID-19_immunity_mortality.

*Competing Interests*

The authors have no conflicts of interest to declare.

*Funding*

This work has been partially funded by the German Research Foundation (DFG) under Grant Nr. KA 1188/13-1, by the German Federal Ministry of Education and Research, and by the Bavarian State Ministry for Science and the Arts.

*Author contributions*

Conceptualization: MW, WH, HK; Data Curation: AB, KK, MW, RS; Formal Analysis: MW, HK; Funding Acquisition: HK; Investigation: WH, MW; Methodology: HK, AB, MW; Project Administration: HK; Resources: HK; Software: MW; Supervision: WH, HK; Validation: MW; Visualization: AB, RS; Writing – Original Draft Preparation: WH, MW; Writing – Review & Editing: MW, WH, AB, KK.

*Acknowledgments*

This work is dedicated to Karl-Walter Jauch, Munich, chairman of the Bavarian Vaccination Committee, on the occasion of his 70$^{th}$ birthday.

**Abbreviations**

| | |
|---|---|
| ARDS | Acute respiratory distress syndrome |
| BayIfSMV | Bayerische Infektionsschutzmaßnahmen-Verordnung |
| CFR | Case fatality rate, |
| CI | Confidence interval |
| CoV-2 | Corona virus type 2 |
| COVID-19 | Coronavirus disease 2019 |
| HR | Hazard ratio |
| ICU | Intensive care unit |
| IfSG | Infektionsschutzgesetz |
| IQR | Interquartile range |
| LGL | Landesamt für Gesundheit und Lebensmittelsicherheit |
| LMU | Ludwig-Maximilians-Universität |
| RAT | Rapid antigen test |
| RT-PCR | Real time polymerase chain reaction |
| SARS | Severe acute respiratory syndrome |
| STIKO | German Standing Commission on Vaccination |

# Figures

Figure 1: Strobe selection

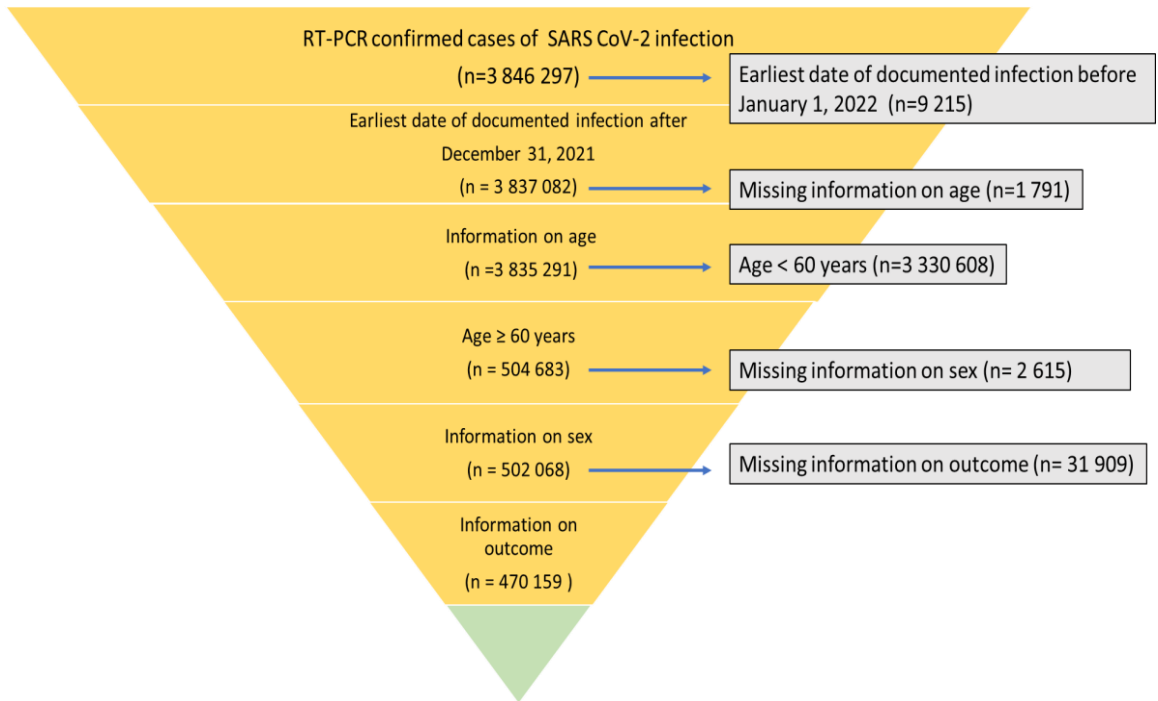

Figure 2: 60-day survival probability (Kaplan-Meier graph) depending on the level of immunity

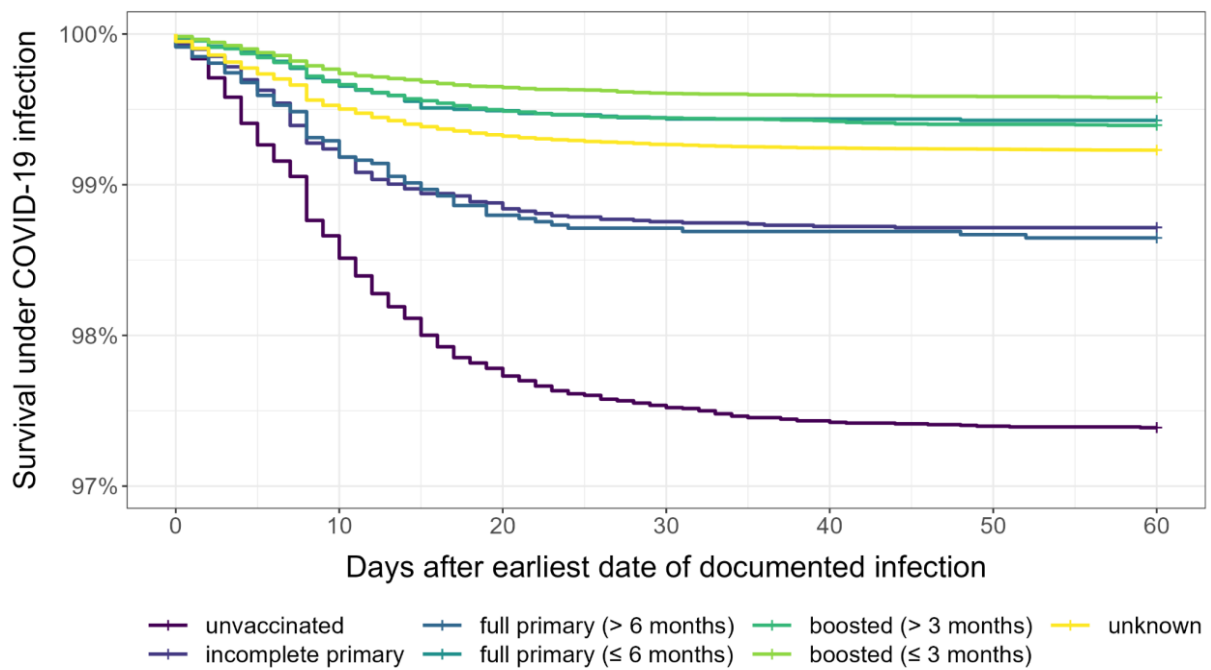

Figure 3: Adjusted association of age (left) and earliest date of documented infection (right) with the risk of death; gray areas indicate 95% confidence bands.

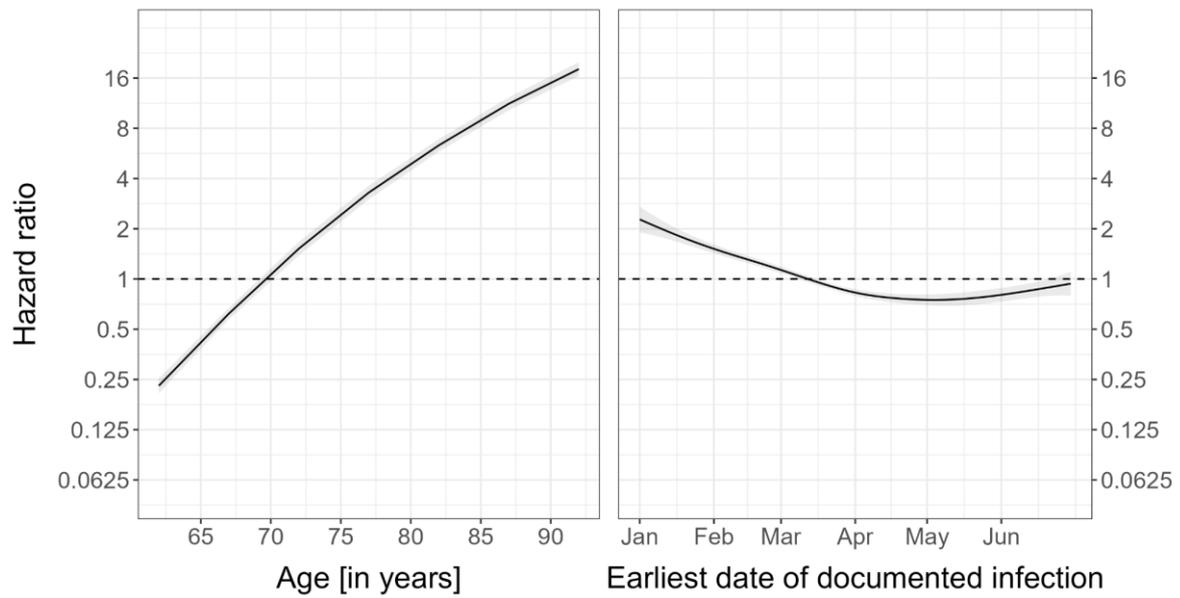

Figure 4: Adjusted association of level of immunity with the risk of death (with 95% confidence intervals). Reference category is unvaccinated cases.

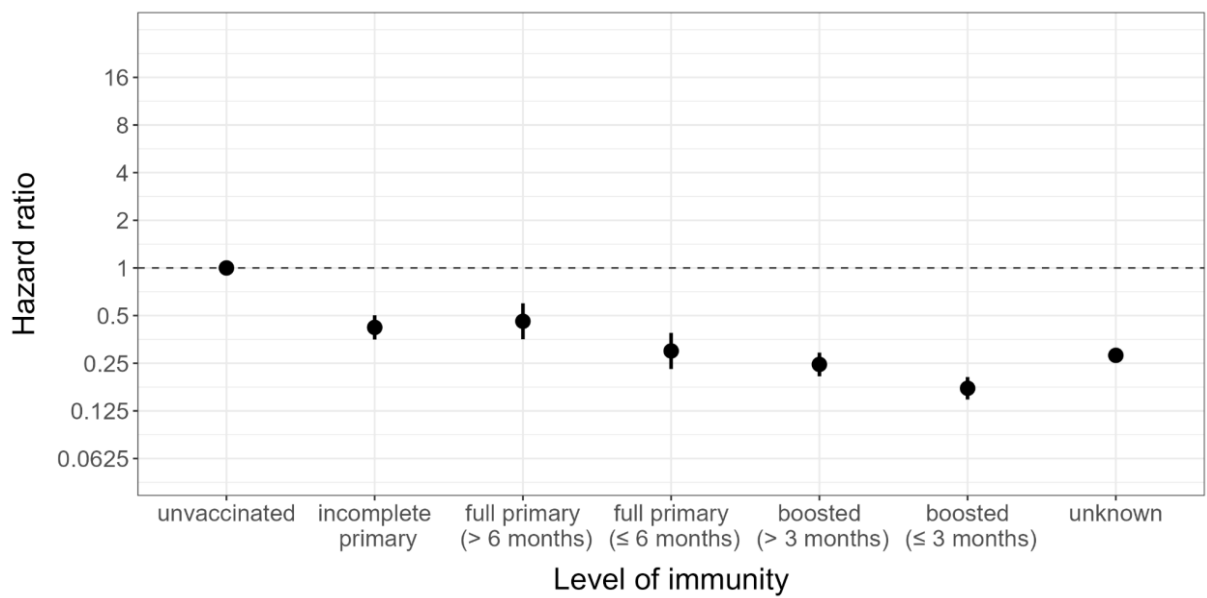

**Tables**

Table 1: Demographic characteristics and outcome of cases of 60 years and older in Bavaria

| Clinical variable | | Covid-19-related death | |
| --- | --- | --- | --- |
| | | Yes | No |
| Sex | Female | 1 851 (0.7%) | 246 461 (99.3%) |
| | Male | 1 985 (0.9%) | 219 814 (99.1%) |
| | Diverse | 0 (0%) | 48 (100%) |
| Age (years) | 60 - 79 | 1 079 (0.3%) | 374 623 (99.7%) |
| | 80+ | 2 757 (2.9%) | 91 700 (97.1%) |
| Level of immunity (LI) | | 511 (2.6%) | 19 050 (97.4%) |
| | Incomplete primary LI | 165 (1.3%) | 12 680 (98.7%) |
| | Full primary LI established more than six months before the earliest date of documented infection | 63 (1.4%) | 4 593 (98.6%) |
| | Full primary LI established less than six months before the earliest date of documented infection | 63 (0.6%) | 10 933 (99.4%) |
| | Boosted LI established more than three months before the earliest date of documented infection | 186 (0.6%) | 30 493 (99.4%) |
| | Boosted LI established less than three months before the earliest date of documented infection | 204 (0.4%) | 48 200 (99.6%) |

# *Online supplement to*

# Association of vaccine-induced or hybrid immunity with COVID-19-related mortality during the Omicron wave - a retrospective observational study in elderly Bavarians

**Introduction**

Infection protection measures and course of the omicron wave

To control the fourth SARS-CoV-2 (Delta) wave in the autumn of 2021, the Bavarian state government declared a disaster situation on November 11, 2021 (**40**). On November 24, the 15th Bavarian Ordinance on Infection Protective Measures (Bayerische Infektionsschutzmaßnahmen-Verordnung, BayIfSMV**)** (**41**) enabled uniform protective measures particularly relevant for unvaccinated people, and clearly less stringent for vaccinated people with booster vaccinations.

The intensified COVID-19 mastering strategy included restrictions for group and public meetings and gatherings, border entry restrictions, isolation of cases, quarantine of close contacts, contact tracing by the health offices, closing of indoor clubs and discos, banning spectators from attending major national events such as soccer matches, and the use of personal protective measures. Only vaccinated or recovered persons had access to leisure and cultural events, gastronomy as well as to services and accommodation close to the body (2G, 2G+ rules). The disease had been largely controlled until the end of 2021 in Bavaria, Germany.

During the subsequent SARS-CoV-2 omicron epidemic, however, infections increased again from a minimum of 183 cases/100000 residents on December 31, 2021, to a peak of 2207 cases/100,000 residents on March 24, 2022 as estimated in nowcasting projections adjusting for potential underreporting of daily infection and hospitalization rates (**42**). With a one- to three-week delay, weekly numbers of hospital and intensive care unit (ICU) admissions also started to grow again. The peaks of hospital admission nowcasts (17.1/100000, almost 400% of the previous minimum) and ICU admissions (2.8/100000, more than 200% of the previous minimum) were observed during the third week of March 2022 (**43-45**).

Simultaneously, however, peaks of weekly hospital and ICU admissions, and of Omicron-associated weakly death toll (350 cases) in Bavaria were 18%, 26% and 55% lower than corresponding maximum rates observed during the preceding Delta surge (**1, 2, 45**). At the same time, occupancy of Bavarian ICU beds with COVID-19 patients rose only by about 43% (from 313 on January 27 to a maximum of 448 on March 22, 2022) reflecting a milder course of disease and, as a consequence, comparably shorter ICU length of stay (**45**). Consequently, on February 16, 2022, the Bavarian state government agreed on the stepwise withdrawal of SARS-CoV-2 mitigation measures over the next weeks until April 3. However, basic protective measures such as the obligation to wear a mask, for example in hospitals or public transportation, remained in place also thereafter.

**Methods**

Classification of causes of death, and calendar date of death

The German Infection Protection Act (Infektionsschutzgesetz, IfSG) obliged physicians to inform the local health authority within one working day of a death associated with a SARS-CoV-2 infection or COVID-19. Further investigations were performed at the local health authorities to clarify the cause of death if needed.

In an exploratory data analysis of all cases older than 59 years in which the date of death followed or was identical to the earliest date of documented infection, we found that the median time elapsing between the earliest date of documented infection and date of death was 8 days. Therefore, if the earliest date of documented infection was later than the date of death (242 cases), we set it to 8 days before the date of death. Vice versa, if the calendar date of death was missing (37 cases), we set this date to 8 days after the earliest date of documented infection. By doing this, we were able to include all recorded mortality cases in the analyses.

Level of immunity

Subjects with an incomplete primary level of immunity were defined as those who had had only one vaccination at least two weeks before the earliest date of documented infection. Subjects with a full primary level of immunity were defined as those who had had a first vaccination (or an earlier SARS CoV-2 infection), and in whom the second (first) vaccination had been administered at least two weeks before the earliest date of documented infection (**46**). Subjects with a single boosted level of immunity must have had a full primary level of immunity, and must have had their third vaccination at least seven days before the earliest date of documented infection (**47**). Irrespective of the type of vaccine, cases who had had a second booster vaccination were counted among those with a single booster immunization.

To study the effectiveness of the level of immunity, we analyzed this level according to the time that had elapsed since the administration of the most recent dose of vaccine (unvaccinated, three months, six months). There was evidence from the Delta wave that protection against COVID-19-related severe disease may only decrease by less than 10% up to six months after the second vaccine dose (**48, 49**). A shorter time interval (three months) was chosen to analyze the effectiveness of a boosted level of immunity; some studies had suggested a faster waning of first or second booster vaccine effect in cases infected by the Omicron variant (**4, 29, 30, 37, 38**).

Vaccines and timing of vaccine authorizations

In Germany, large-scale COVID-19 vaccination started in January 2021. On October 18, the German Standing Commission on Vaccination (STIKO) recommended a booster vaccination with an mRNA vaccine (at the earliest six months after the second vaccination). The recommendations were later extended to other vaccines and different vaccination schedules. Candidates were recovered patients and high-risk persons (immune deficiencies, over 70 years of age, residents cared for in institutions for the elderly, staff in medical and nursing facilities), and all persons with a preceding Jcovden® vaccination. On November 29, recommendations were extended to people older than 18 years, and on December 21, to all persons with a shortened interval of at least three months. On December 13, a modification of the IfSG had already implemented a "facility-related" vaccination obligation applying to employees in clinics, nursing homes, outpatient care services and similar facilities. Finally, on February 17, 2022 STIKO published a new recommendation: according to this, high-risk patients should receive a second booster vaccination with the previous mRNA (at the earliest three months after the first booster vaccination) (**47**).

Of vaccine doses delivered until the end of calendar week #26, Cominarty® (Pfizer-BioNTech; BNT162b2) had been the most popular vaccine given to qualifying individuals (91% of delivered), followed by Vaxzevria® (Oxford-AstraZeneca; ChAdOx1) (89%), Spikevax® (Moderna; mRNA-1273) (87%), Jcovden® (Janssen-Cilag/Johnson and Johnson, Ad26.COV2.S) (69%) and Nuvaxovid® (Novavax , NVX-CoV2373) (8%) (**47**).

Statistical analysis

Hazard ratios describing risks of death were estimated using Cox proportional hazard models (**16**) with penalized splines (**17**). Since an explorative data analysis revealed that about 99% of deaths had occurred in cases during the first 60 days after the earliest date of documented infection, we restricted our analysis to this time window of follow-up; thereby, we also reduced the risk of bias caused by violations of the proportional hazard assumption.

Adjusted absolute risk reduction for cases with a certain level of immunity compared to unvaccinated cases was calculated as the difference in mean predicted event probability up to 60 days, if all observed cases were unvaccinated and had the considered level of immunity, respectively (**50**). Effectiveness as a measure of relative risk reduction (**51**) was estimated using the same strategy. Confidence intervals for both measures of risk reduction were based on a non-parametric bootstrap with 1,000 bootstrap samples (**50**).

We constructed a baseline model adjusting for sex, age category, level of immunity and the earliest date of documented infection. We conducted a subgroup analysis in cases older than 79 years, and three sensitivity analyses to test the robustness of estimates for SARS-CoV-2 infection-related mortality:

   a) by analyzing 30-day mortality, since there is evidence that a longer follow-up may be affected by violations of the proportional hazard assumption (**52**).
   b) by using a modified definition of the dependent variable only considering cases for which physicians had certified COVID-19 as the leading cause of death. Cases in which a SARS-CoV-2 infection had not been the leading cause of death, or cases of which cause of death had been unidentifiable or had not been determined, were treated as being censored at the date of death.
   c) by restricting our analysis to subjects whose level of immunity had been known

Cox regression models were estimated with the coxph function from the R package survival (version 3.4-0) (**53**). The R package pec (version 2022.05.04) (**54**) was used to predict event probabilities up to 60 days.

**Results**

Primary analysis

Figure S1: Frequency of weekly cases of which outcome was unknown

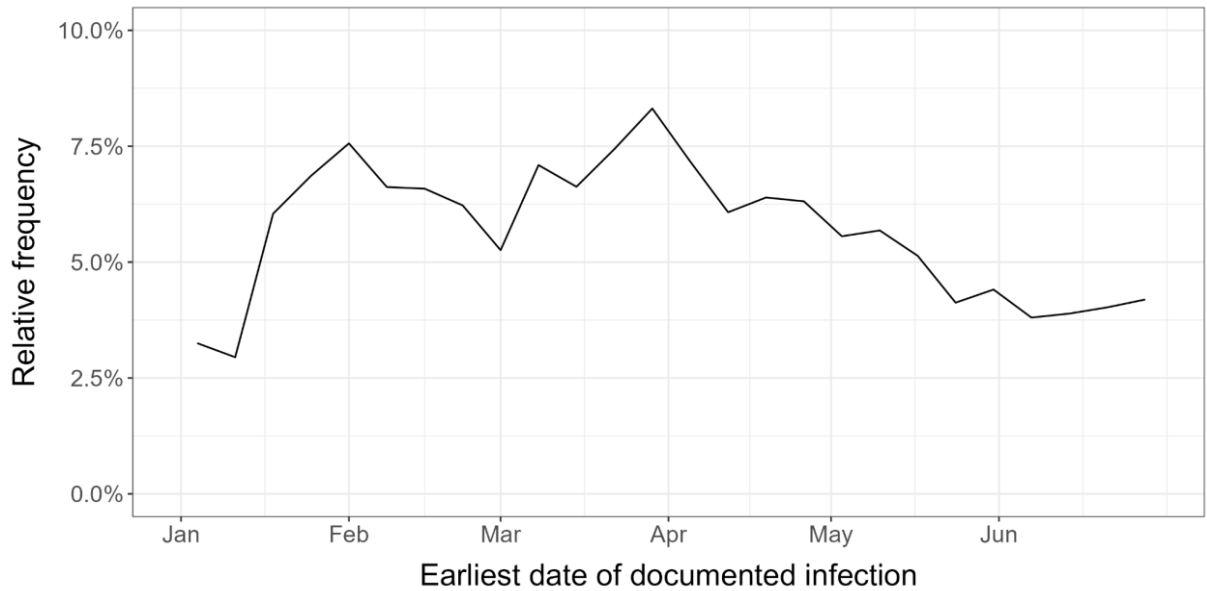

Figure S2: Frequency of cases according to cause of COVID-19 related death

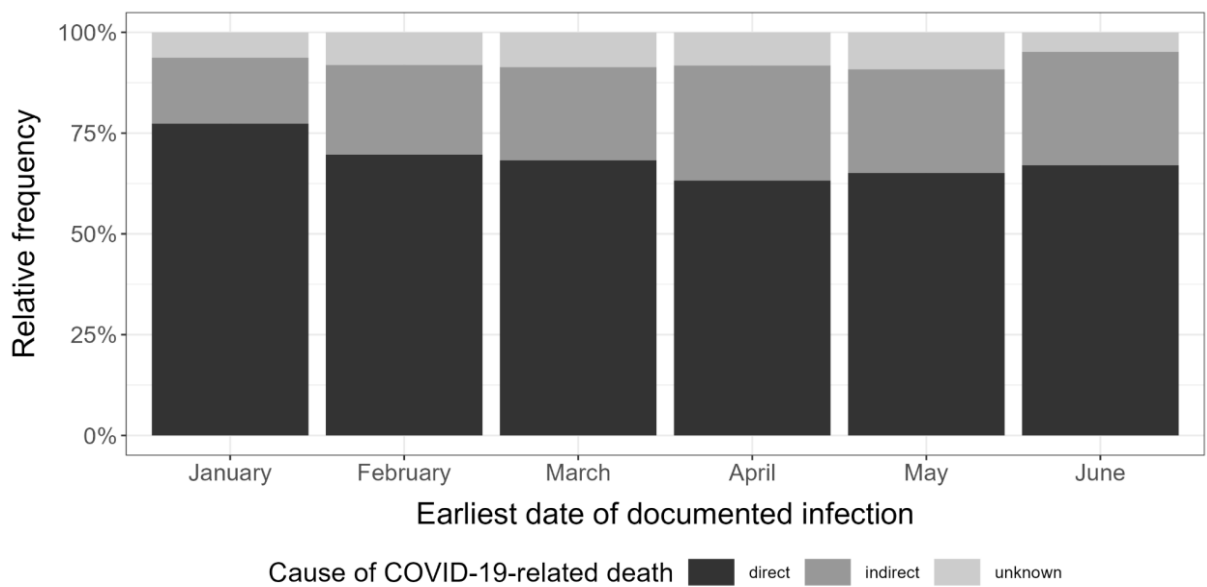

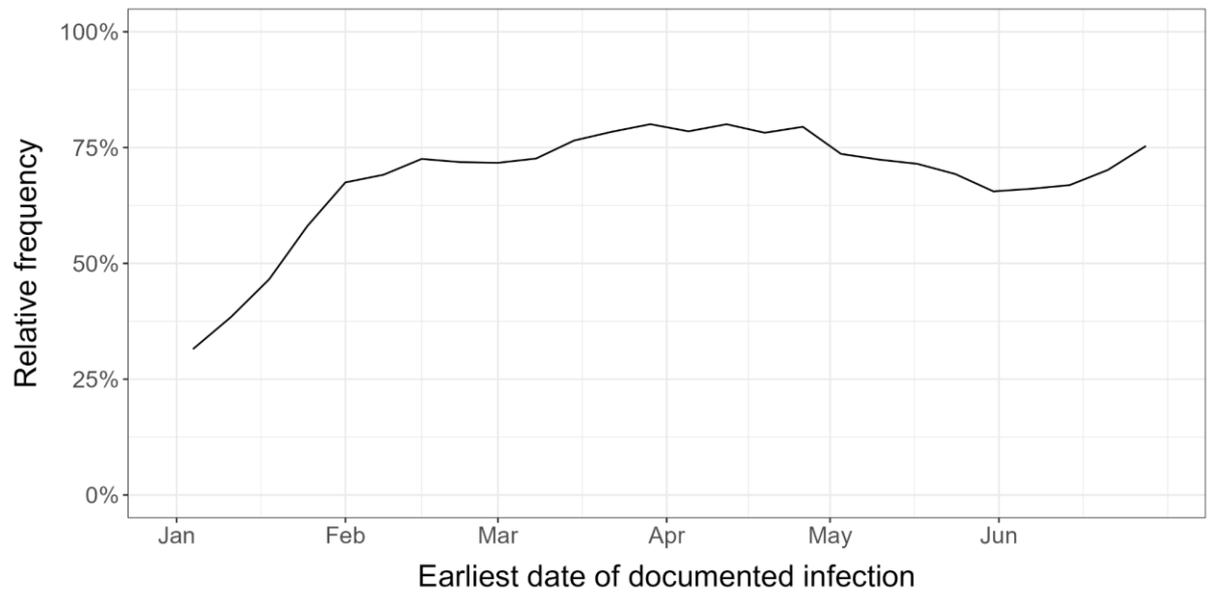

Figure S3: Frequency of weekly cases of which the level of immunity was unknown

Subgroup analysis in cases older than 79 years

We analyzed 94 457 cases. With a full primary level of immunity, HR increased after six months from 0.33 (95% CI 0.24- 0.46) to 0.53 (95% CI 0.40-0.71), effectiveness decreased from 65.5% (95% CI 53.6-75.4%) to 45.4% (95% CI 29.8-59.3%), and absolute risk reduction from 5.8% (95% CI 4.5-7.0%) to 4.0% (95% CI 2.6-5.4%), respectively. With a boosted level of immunity of which the associated HR increased after three months from 0.21 (95% CI 0.17-0.26) to 0.27 (95% CI 0.22-0.33), effectiveness decreased from 77.8% (95% CI 73.5.-81.8%) to 71.8% (95% CI 66.0-77.0%), and absolute risk reduction from 6.9% (95% CI 6.0-7.8%) to 6.4% (95% CI 5.4-7.3%).

Figure S4: Adjusted association of age (left) and the earliest date of documented infection (right) with the risk of death; gray areas indicate 95% confidence bands.

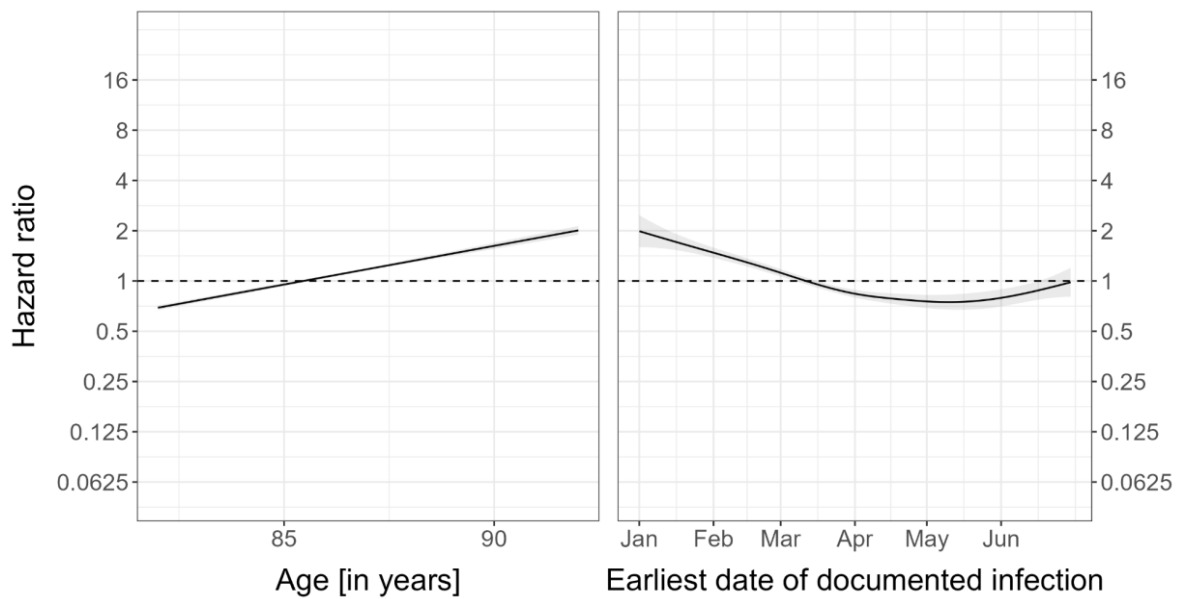

Figure S5: Adjusted association of the level of immunity with the risk of death (with 95% confidence intervals). Reference category is unvaccinated cases.

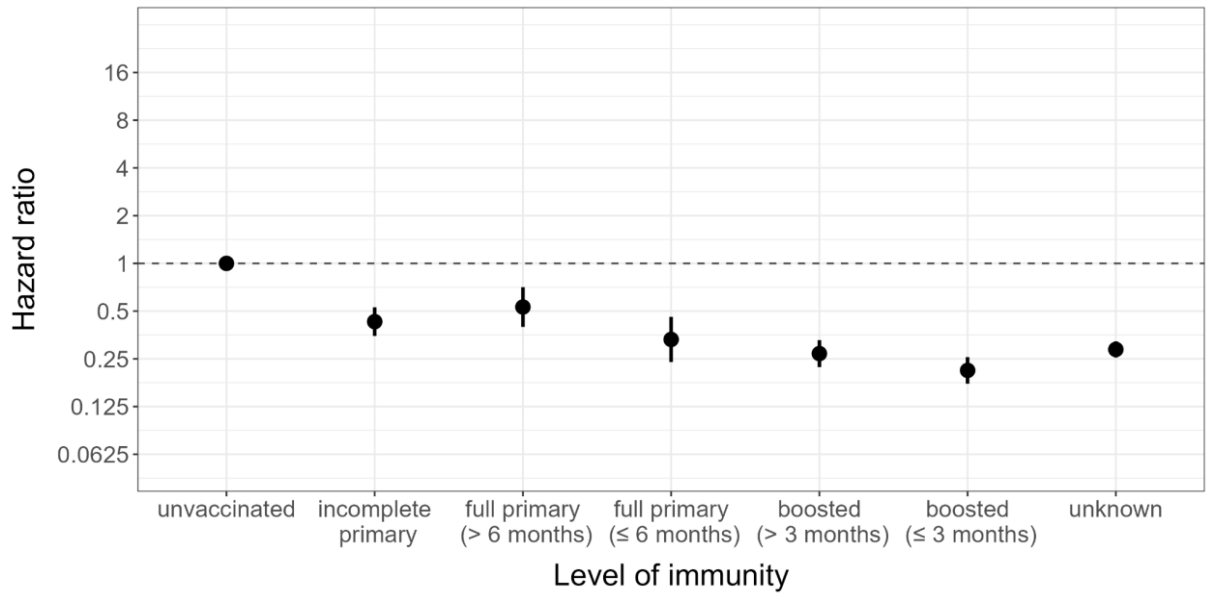

Sensitivity analyses

Sensitivity analyses, which only considered cases which had died before day 30 after the earliest date of documented infection, in which leading cause of death had been COVID-19, or for which the level of immunity had been known, yielded results qualitatively similar to the (Supplementary Figures 6 to 11). Quantitatively, however, HRs for associations of level of immunity with outcome were higher when only analyzing patients in whom COVID-19 had been the leading cause of death. Furthermore, compared to the main analysis, adjusted HRs were lower at the beginning of 2022 in those cases of which the level of immunity had been known precisely (see time effect in Supplementary Figure S10).

*a) 30-day mortality as endpoint*

Figure S6: Adjusted association of age (left) and the earliest date of documented infection (right) with the risk of death; gray areas indicate 95% confidence bands.

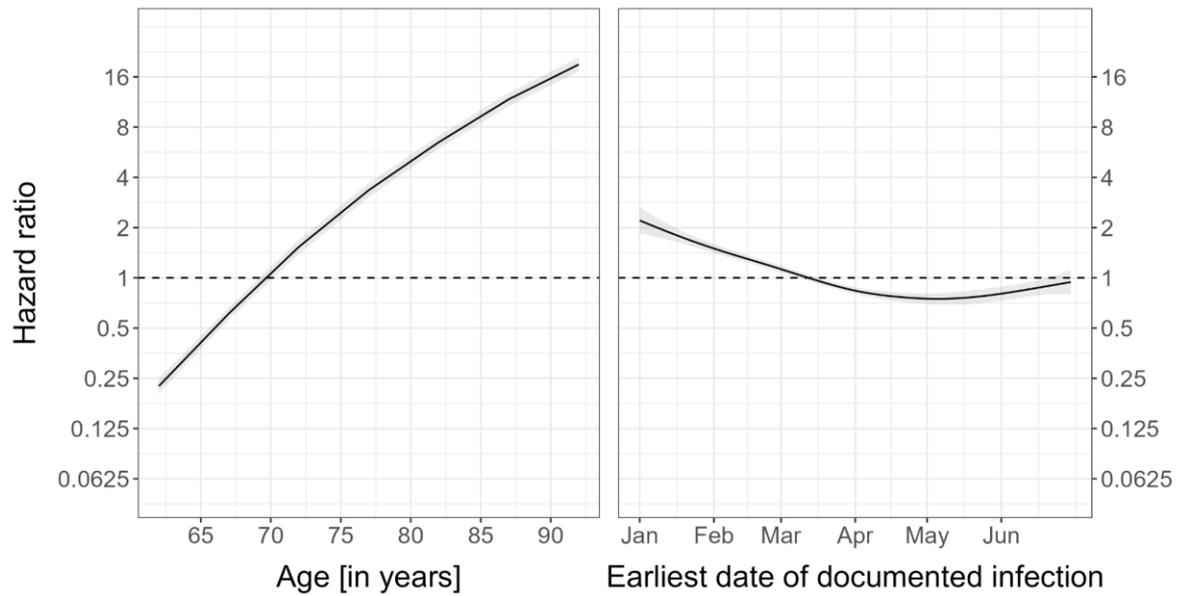

Figure S7: Adjusted association of the level of immunity with the risk of death (with 95% confidence intervals). Reference category is unvaccinated cases.

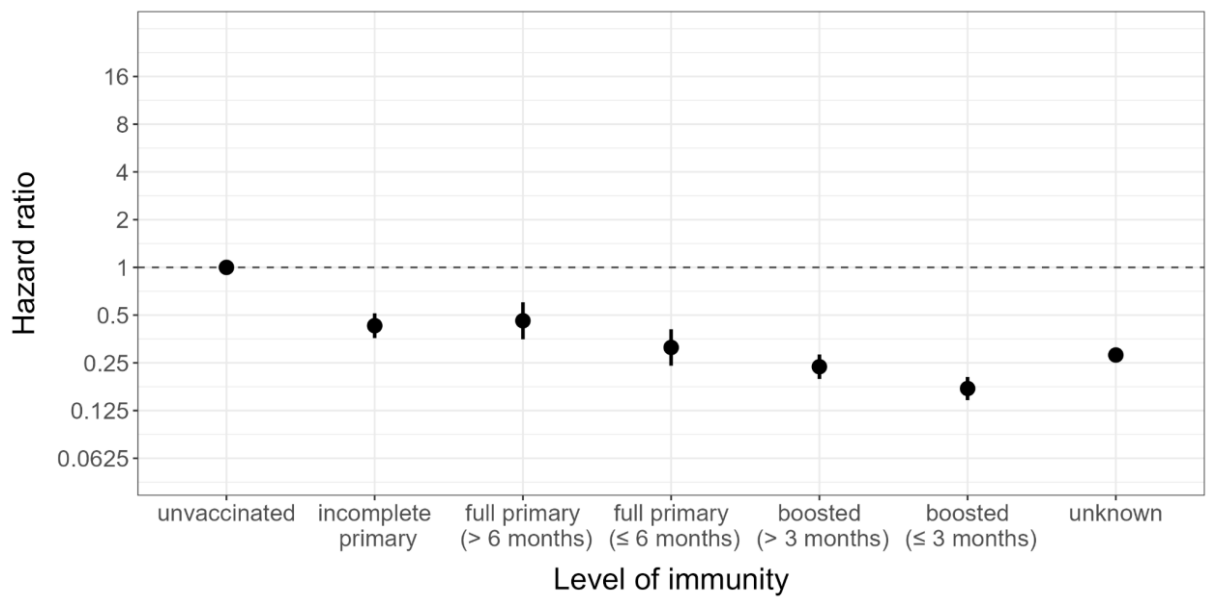

b) *Modification of the dependent variable (cases in which COVID-19 was the leading cause of death)*

Figure S8: Adjusted association of age (left) and the earliest date of documented infection (right) with the risk of death; gray areas indicate 95% confidence bands.

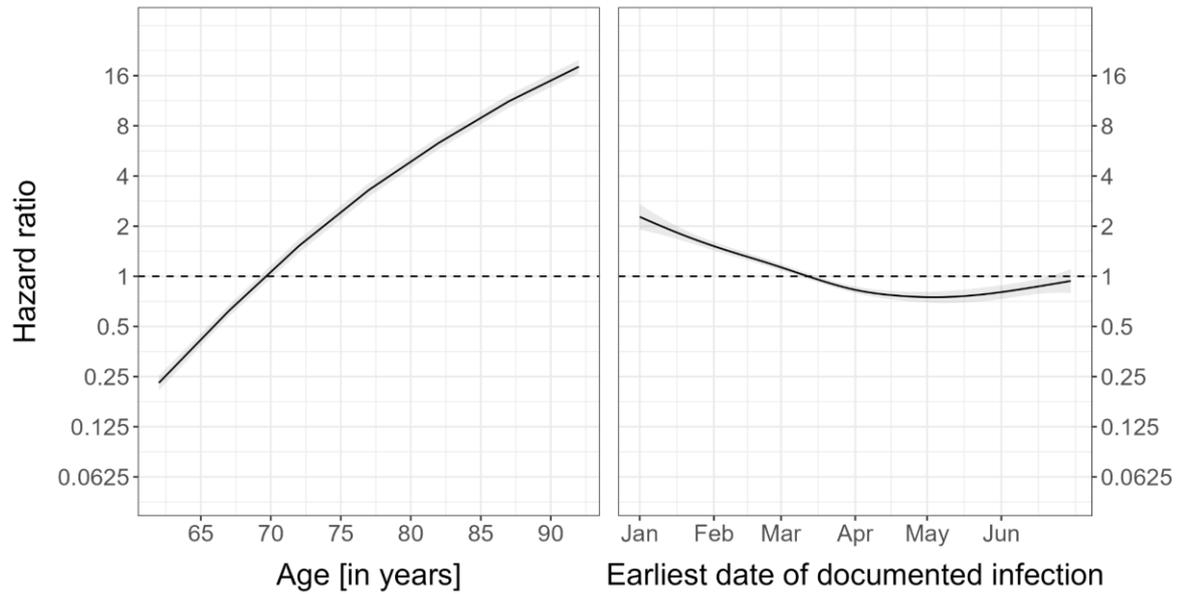

Figure S9: Adjusted association of the level of immunity with the risk of death (with 95% confidence intervals). Reference category is unvaccinated cases.

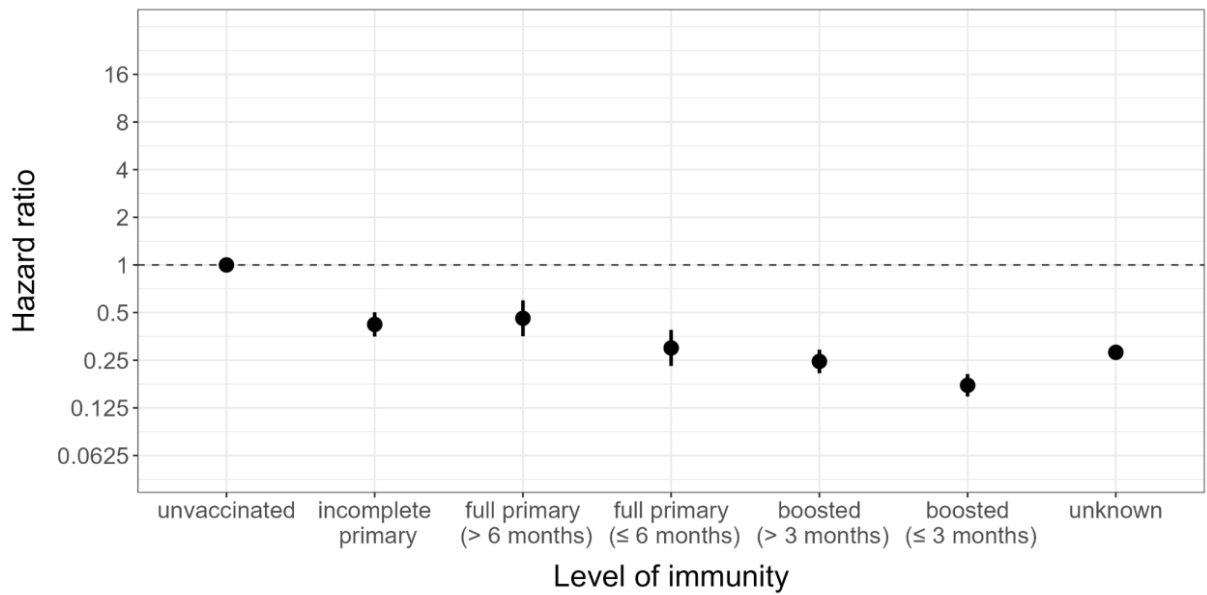

*c) Cases of which the level of immunity had been known*

Figure S10: Adjusted association of age (left) and the earliest date of documented infection (right) with the risk of death; gray areas indicate 95% confidence bands.

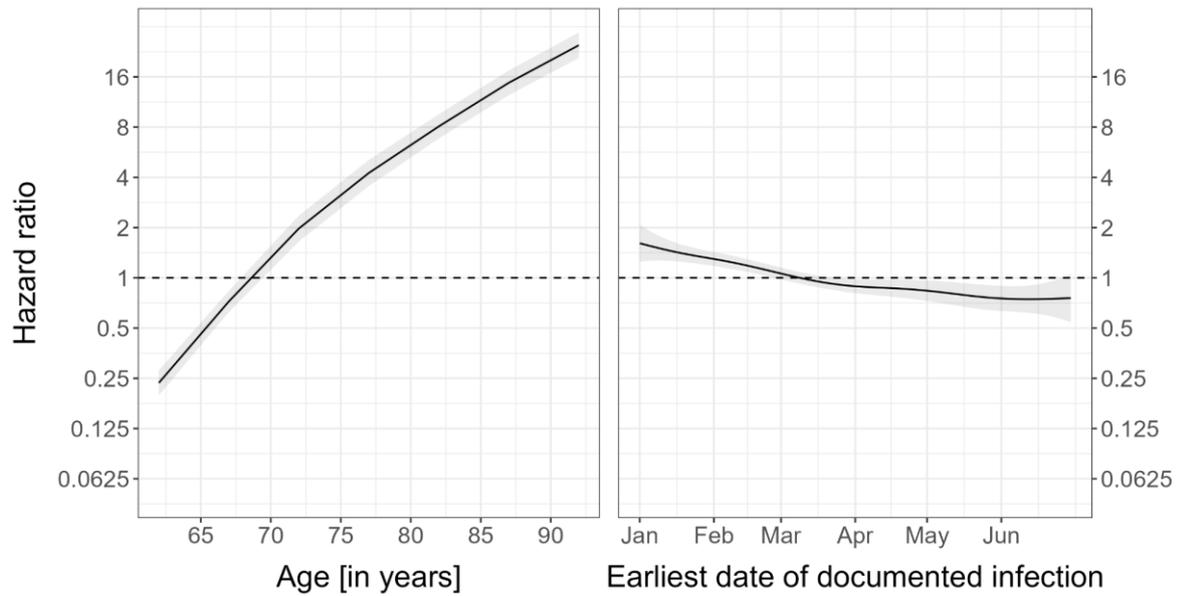

Figure S11: Adjusted association of the level of immunity with the risk of death (with 95% confidence intervals). Reference category is unvaccinated cases.

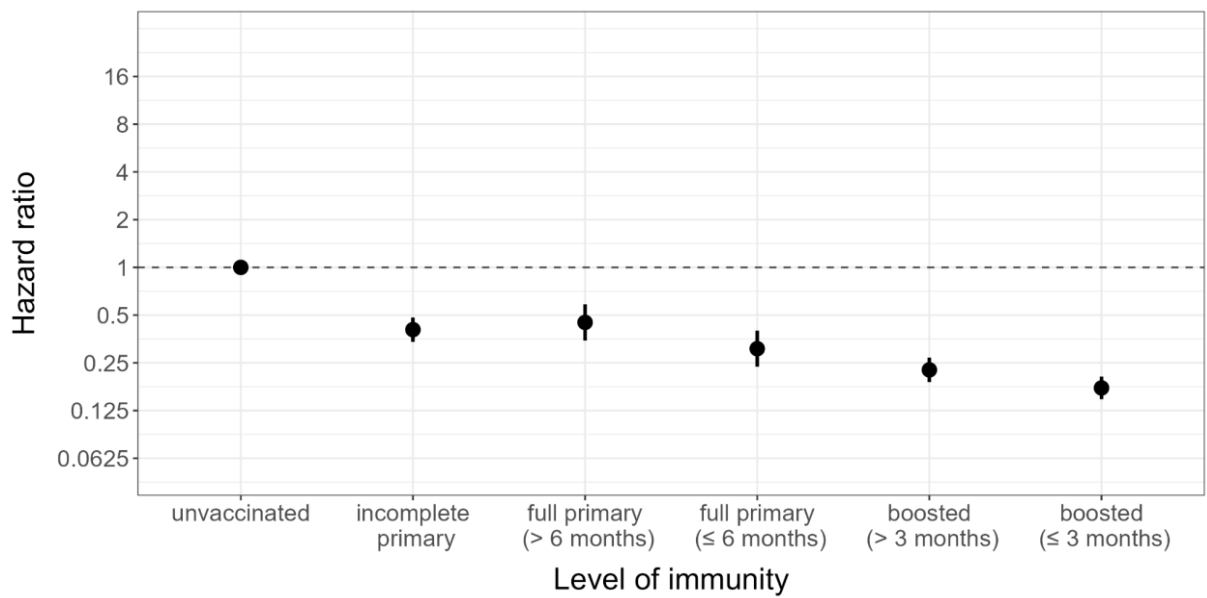

**Discussion**

Mortality

Despite vaccination efforts (**55**) and an increasing SARS-CoV-2 infection rate (**56**), which both should enhance immunity on a population level, our population-based cohort analysis provides evidence that in Bavarian cases older than 59 years, 60-day risk of death from any cause was still relevant during the first omicron wave (CFR 0.83%). The time elapsing between the earliest date of documented infection and death was four to 14 days in most cases (median eight days). This time span was shorter than that found in a recent German autopsy study (median two weeks, interquartile range 2 to 6 weeks) (**57**) which, however, also included younger patients having a better tolerance against severe COVID-19; furthermore, the definition of the date of documented infection was different, and autopsies were only done in cases which had died in the pre-Omicron period.

COVID-19 as cause of death

In our patients, COVID-19 was reported to be the leading cause of death in about 70% of cases. This frequency was lower than rates observed by the German Federal Statistical Office (Destatis) (83%), which analyzed 36 291 cause of death certificates issued for patients infected by SARS-CoV-2 (**58**). Postmortem examinations of hospitalized patients in Germany and in other countries found a rate of 75-86% (**57, 59, 60**). Exclusive examination of hospitalized patients, however, may have caused a selection bias explaining the higher rates from autopsy studies. Furthermore, COVID-19 may have been more often the leading cause of death in patients infected by the Alpha- and Delta-variant than by the Omicron-variant (**60**). The most common clinical sequelae by which an acute SARS CoV-2 infection may directly cause death, include ARDS, pneumonia-induced respiratory and multi-organ failure, and shock/acute cardiac injury (**61**).

Demographic predictors of mortality

Independent from the earliest date of documented infection and level of immunity, our study confirmed established predictors of mortality in SARS-CoV-2 infected patients. Well-known risk factors are male sex (**62**) and older age (Figure 3). The risk associated with age may be attributed to age-associated comorbidities (**63**), but also to intrinsic age-related quantitative and qualitative changes in the immune system. In the elderly, an altered immune system may not only increase the susceptibility to and defense against infections, but also decrease the response to therapeutics and to vaccines (**64**).

Association of the earliest date of documented infection with the risk of death

We identified a significant change of mortality risk during the period under study (from the beginning of January 2022 until the end of June). This non-linear change was adjusted for age, sex and level of immunity. Hazard rates for a certain calendar day of documented infection decreased steadily reaching a minimum in early May (about one month and a half after the peak of the Omicron BA.1/BA.2 wave). Subsequently, however, and running parallel to the upcoming Omicron BA.5 wave, hazard rates increased again (Figure 3). Several mechanisms might be considered to explain this observation:

a) Comorbidity

It might appear tempting to assume that the initial decrease of HR for death over time reflected a progressive virus spreading into a less vulnerable part of the population characterized by a younger age or less age-associated comorbidities. Over the course of the Omicron pandemic in Germany, however, younger age groups were infected before the virus reached older populations. For age category 60 – 64 years, weekly numbers of reported SARS-CoV-2 infections surmounted 700/100,000 residents in the fifth registration week of 2022, whereas in age category 80-85 years, this incidence was only reached in registration week #11 (**65**). Furthermore, since we adjusted changes of HR over time for age category, and since multimorbidity significantly increases with age (**66**), it is unlikely that a varying age-related comorbidity was responsible for the time-dependent decrease of HR for death.

b) Omicron variant

During the period under study, the Omicron variants BA.1, BA.2 and finally BA.5 progressively replaced the Delta variant B.1.617.2., which had been the leading cause of infection during autumn of 2021 in Bavaria. The frequencies of Omicron BA.1 and BA.2 variants in Bavaria were presumably > 95% since the third and fourteenth calendar week 2022, respectively (**12**). Since calendar week #22, however, more than 50% of SARS-CoV-2 infections were most likely caused by the BA.5 variant.

Recent studies suggested that hospitalization and mortality rates might be lower for infections with the Omicron variant and sublines than for the Delta variant, even after controlling for confounders such as patient demographics, previous infection, and vaccination status (**33, 67-71**). However, it is uncertain whether this observation may explain continuously declining hazard rates until the end of April. The reduction in the risk of death for Omicron compared with Delta may be less pronounced in the elderly (**67**), and the precise mechanisms of this attenuated risk of death are still unclear (**72**). Furthermore, risk of severe outcomes, and vaccine effectiveness against severe disease is presumably similar for the BA.1 and BA.2 variant (**71**, **73).**

The new increase of hazard rates after mid-May was running parallel to the upcoming Omicron BA.5 wave. It is currently controversial whether disease severity or vaccination effectiveness differs significantly between the Omicron sublinages BA.1/BA.2 and BA.5. Two recent studies found that the risk of severe hospitalization/death was similar in the BA.4/BA.5 and BA.1 waves (**74, 75**), whereas a study from Denmark provided some evidence that BA.5 infections were associated with an increased risk of severe disease compared with BA.2 infections (**22**). Furthermore, protection provided by a preceding infection is only modest against BA.4 or BA.5 reinfection when the previous infection had been caused by a pre-omicron variant (**76**). There are also divergent results concerning vaccine effectiveness and associated waning over time (**29, 39, 77**). Our findings are consistent with the hypothesis that BA.5 infections are somewhat more dangerous than preceding Omicron variants.

c) Medication

On February 25, 2022 BfArM officially approved Paxlovid® (Nirmatrelvir/Ritonavir) as an additional first line therapy of COVID-19 (**78**). In the same month, several German medical societies had issued specific guidelines recommending that Paxlovid® may be given to adult high-risk in- and outpatients during the early stage of COVID-19 (**79, 80**). Nirmatrelvir/Ritonavir was found to effectively reduce rates of hospitalization and death due to Omicron infection among patients 65 years of age or older (**81).** It has been estimated, however, that only 1500 doses were prescribed in Bavaria until the end of June (**82**) rendering this drug an unlikely candidate for exclusively explaining temporal changes of risk of death.

d) Time of the year

It is unlikely that a seasonal influenza-like illness or extreme weather conditions (periods of great cold) were responsible for the high risk of death in early 2022 (**83, 84**). Seasonal weather effects, however, cannot be excluded. Transition to warmer weather is associated with less air pollution and higher humidity. Both conditions were found to be independently associated with lower SARS-CoV-2 infection rates, preferentially in high risk subjects suffering from an impaired nasopharyngeal mucosal immune defense (**85-87**).

d) Positivity rate

Between March and June, changes of HR over time were running parallel to changes of positivity rate (the percentage of all coronavirus tests performed that are actually positive). Positivity rate regarding the whole population peaked at the middle of March, and, again, at the end of the registration period (**42**). Since a greater positivity rate was found to be associated with a higher number of unrecognized infections and a greater case fatality rate (**88, 89**), it is possible that detection and therapy of infections in high risk patients may have been delayed in March and June increasing mortality risk at that time.

**Limitations**

Type of vaccine

We were unable to stratify our results according to the type of vaccine used. However, with regard to vaccine effectiveness against COVID-19 hospitalization and death due to Omicron infections, there may be no discernable differences in effectiveness of BNT162b2 vaccine versus mRNA-1273 vaccine (**90). However,** rates of COVID-19-related death were found to be consistently lower among people fully vaccinated with BNT162b2 than with ChAdOx1 (**48**) or Ad26.COV2 (**34**).

Cause of death

There was a certain risk of misclassification /underreporting of death from COVID-19 by practitioners who were not necessarily the attending physician. If death had occurred in the domestic environment, such as private homes or non-hospital care facilities, post-mortem external examinations were usually performed at the place of death by on-duty general practitioners or emergency physicians who were sometimes not aware of the underlying disease (**91**). This unawareness may have led to incorrect cause of death certifications or misclassifications. Frequency of misclassifications, however, is presumably rare, affecting less than 10% of deaths (**92-94**).

To account for this weakness we performed a sensitivity analysis in which we only included cases in which COVID-19 had been the leading cause of death. Since this sensitivity analysis yielded results qualitatively similar to that of the main analysis (see Figures S6 and S7), it is unlikely that misclassifications interfered with our results.

Missing information on level of immunity

For the majority of our cases, we did not have information on the level of immunity. By assigning a separate level of immunity to these cases, we were able to adjust results on the association between known levels of immunity and risk of death to these unknown levels. To further account for this weakness we performed a sensitivity analysis in which we had only included cases of which the level of immunity had been known. Since this sensitivity analysis yielded results qualitatively similar to that of the main analysis (see Figures S8 and S9), it is unlikely that missing information limited the validity of our results.

Vaccinations

We cannot exclude the possibility that the apparent waning of boosted immunity effects was actually due to an uneven distribution of a second booster vaccination between the different boosted levels of immunity. Cases with a double booster were counted among those with a single booster vaccination, and at the end of June 2022, rates of double booster vaccinations had increased up to about 17% in the elderly Bavarian population (**47**). A fourth dose of a COVID-19 vaccine, administered during the Omicron era, may further reduce risk of death from all causes also in the elderly (**31, 34, 38, 95**).

Comorbidities

a) Numerous comorbidities such as chronic heart failure, chronic obstructive pulmonary disease, diabetes, hypertension and stroke are associated with a higher mortality risk even in vaccinated adults (**25, 96**). Cancer or other diseases associated with immunosuppression may be particularly important, as those conditions have been associated with reduced post-vaccination seroconversion, antibody responses and vaccine effectiveness, and with an extraordinarily high risk for death (**7, 14, 97**). However, among predictors for the risk of COVID-19 death, comorbidities are far less important than age, especially after booster vaccinations (**14**). Furthermore, it has been repeatedly shown that adjustments to pre-existing health conditions will only quantitatively, but not qualitatively change hazard ratios describing associations between variables such as age or SARS-CoV-2 variants, and severe outcomes including death (**33, 67**). Since multimorbidity significantly increases with age (**66**), and since we adjusted our results for age, it is likely that at least a certain portion of chronic comorbidities had been considered by our analysis.

b) Absence of comorbidities may lead to healthy vaccinee bias, by which vaccine recipients are healthier than their unvaccinated peers, thereby possibly inflating the hazard ratios calculated for higher levels of immunity (**98**). A recent analysis, however, suggested that the magnitude of overestimation by healthy vaccinee bias is unlikely to be substantial (**27**).

c) The staggered timing of vaccine authorizations by the German Standing Commission on Vaccination (STIKO), propagating earlier rollout in high-risk populations, might result in spuriously lower effectiveness of higher levels of immunity. Instead to waning immunity, this finding may have been also due to higher risk of patients first prioritized for COVID-19 vaccination (**99**).

Health behavior, socio-economic status, minority ethnic populations

Potential confounders include key worker status, deprivation rank, university degree, household deprivation or country of birth. Thus, some minority ethnic populations may have excess risks of adverse COVID-19 outcomes compared with the White population (**100**). However, according to recent studies, adjusting for sociodemographic characteristics or ethnicity had only a small effect on the association of e.g., SARS-CoV-2 variants with death (**14, 67**).

BMI

Individuals at the two extremes of BMI distribution (very low and very high BMI) may be at greater risk of death from COVID-19 than individuals of normal weight (**7, 101**). This association, however, may not apply to all situations. For example, after the third vaccination dose, some studies found little evidence of an association of BMI with severe outcomes (**102, 103**).

Testing behavior

Many infected cases may have remained undetected because of a mild or asymptomatic infection, lack of testing, or increasing use of unreported SARS-CoV-2 rapid antigen test (RAT) (**104**). It is also unknown how many study participants had undergone voluntary, mandatory or routine RT-PCR testing. After February 12 2022, RT-PCR testing was only free-of-charge if a preceding RAT or pooled testing of samples by RT-PCR had revealed a positive result (**105**). Therefore, it is highly likely that the majority of our cases had been reported on the basis of a preceding positive RAT. Independent from comorbidities, however, individuals who had voluntary RT-PCR tests after this date might have had more severe symptoms (and possibly, a higher risk for a more severe COVID-19) than people who got tested more routinely, did not seek a confirmation of their RAT result, or did not get tested at all. This difference could bias the estimates of risk of COVID-19 death among individuals who tested positive (**67**).